\documentclass[prd,twocolumn,floatfix,amsmath,nofootinbib,amssymb,floatfix]{revtex4}
\usepackage{graphicx,color,dcolumn,booktabs,bm,multirow}
\usepackage{longtable,lscape}
\usepackage{txfonts}
\usepackage{overpic}
\usepackage{amssymb}
\usepackage{array}
\usepackage{indentfirst}
\usepackage{feynmf}   %{feynmp}
\usepackage{bbding}
\usepackage{slashed}  %for Feynman symbols
\usepackage{cases}
\usepackage{color}
\usepackage{multirow}
\usepackage{epstopdf}
\usepackage{tabularx}
\usepackage{graphicx,color,dcolumn,booktabs,bm}
\usepackage[compat=1.1.0]{tikz-feynman}
\usepackage[colorlinks,
citecolor=blue,
anchorcolor=red,
menucolor=red,
linkcolor=red,
filecolor=red,
runcolor=red,
urlcolor=blue,
frenchlinks=red]{hyperref}
\usepackage{float}

\begin{document}
\title{$\Lambda_{c}(2910)$ and $\Lambda_{c}(2940)$ productions in $p \bar{p}$ annihilation}
\author{Quan-Yun Guo$^{1}$}
\author{Dian-Yong Chen$^{1,2}$\footnote{Corresponding author}}\email{chendy@seu.edu.cn}
\affiliation{$^1$ School of Physics, Southeast University, Nanjing 210094, People's Republic of China}
\affiliation{$^2$ Lanzhou Center for Theoretical Physics, Lanzhou University, Lanzhou 730000, China}
\date{\today}

\begin{abstract}

In this work, we investigate the productions of $\Lambda_{c}(2910)$ and $\Lambda_{c}(2940)$ in the $p \bar{p} \rightarrow \bar{\Lambda}_{c} D^{0} p$ process by utilizing an effective Lagrangian approach, where both $\Lambda_{c}(2910)$ and $\Lambda_{c}(2940)$ are considered as $D^{\ast}N$ molecular states with $J^{P}=1/2^{-}$ and $3/2^{-}$, respectively. In addition to the $t$-channel $D$ and $D^{\ast}$ exchanges, the contribution from light-meson exchange is also considered. At $\sqrt{s}=10$ $\mathrm{GeV}$, our estimations indicate that the total cross sections for $p \bar{p} \rightarrow \bar{\Lambda}_{c} D^{0} p$ are $(130.7^{+397.1}_{-103.9})$ nb, where the central value is estimated with $\Lambda_{r}=1.1$ GeV, and the uncertainties are resulted from the variation of parameter $\Lambda_{r}$ from 1.0 GeV to 1.2 GeV. Our results indicate that the light meson exchange diagrams are very important, which provide a very large smooth background. Moreover, the estimations of the $D^{0}p$ invariant mass spectrum reveal that the peak structure between 2.9 and 3.0 GeV primarily originates from $\Lambda_c(2910)$, while the signal of $\Lambda_{c}(2940)$ is about one order smaller than that of $\Lambda_{c}(2910)$. Furthermore, the resulting Dalitz plot is estimated with $\sqrt{s}=8$ GeV. It is expected that our estimations in the present work can be tested by future experiments at $\mathrm{\bar{P}ANDA}$.

\end{abstract}
	
\maketitle
	%\end{CJK}

\section{Introduction}
\label{sec:Introduction}
%%%%%%%%%%%%%%%%%%%%%%
\begin{figure*}[htb]
     \includegraphics[width=165mm]{feyn.pdf}
     \caption{Diagrams contributing to the production of $\Lambda_c(2910)/\Lambda_c(2940)$ in the $p \bar{p} \rightarrow \bar{\Lambda}_{c} D^{0} p$ process. Diagrams (a) and (b) correspond to the $t$-channel light-meson and $D$ and $D^{\ast}$ exchanges, respectively. Here $\Lambda_{c}$ and $\Lambda^{\ast}_{c}$ refer to $\Lambda_{c}(2286)$ and $\Lambda_{c}(2910)/\Lambda_{c}(2940)$ states, respectively.}
     \label{Fig.1}
\end{figure*}
%%%%%%%%%%%%%%%%%%%%

Since the observation of $X(3872)$ by the Belle Collaboration in 2003~\cite{Belle:2003nnu}, some potential exotic state candidates have been observed and identified by experimental Collaborations including Belle/Belle II, BESIII, and LHCb (See Refs.~\cite{BaBar:2005hhc,Belle:2005lik,Belle:2007umv,BESIII:2013ouc,BESIII:2013qmu,LHCb:2016axx,Belle:2017egg,LHCb:2020tqd,LHCb:2021vvq,Dong:2017gaw,Liu:2019zoy} for examples). Among the possible pentaquark states, $P_{c}(4380)$ and $P_{c}(4450)$ were first observed in the $J/\psi p$ invariant mass distributions of the $\Lambda_{b} \rightarrow J/\psi K^{-}p$ process by the LHCb Collaboration in 2015~\cite{LHCb:2015yax,LHCb:2016ztz}. Later, the same process was analyzed with the dataset collected by the LHCb Collaboration in Run I and Run II in 2019~\cite{LHCb:2019kea}, their results indicated that $P_{c}(4380)$ could be described by the background and a new narrow state, $P_{c}(4312)$ with a statistical significance of 7.3 $\sigma$. Moreover, $P_{c}(4450)$ splitted into two narrow states $P_{c}(4440)$ and $P_{c}(4457)$. In addition to the above observations, the strange counterparts of the $P_{c}$ states, $P_{cs}$, were further observed by the LHCb Collaboration, which included $P_{cs}(4459)$~\cite{LHCb:2020jpq} and $P_{cs}(4338)$~\cite{LHCb:2022ogu}. Due to the fact that the masses of $P_{c}$ and $P_{cs}$ states are close to the thresholds of $D^{(\ast)} \Sigma_{c}$ and $D^{(\ast)} \Xi_{c}$, respectively, the properties of these states have been comprehensively investigated in the molecular frame~\cite{Chen:2019bip,Chen:2019asm,Guo:2019fdo,Liu:2019tjn,Xiao:2019mvs,Xiao:2019aya,Zhang:2019xtu,Wang:2019hyc,Xu:2019zme,Burns:2019iih,Lin:2019qiv,He:2019rva,Du:2019pij,Wang:2019spc,Xu:2020gjl,Xu:2020flp,Ling:2021lmq,Chen:2021cfl,Du:2021bgb,Chen:2022onm,Chen:2020kco,Zhu:2021lhd,Xiao:2021rgp,Wu:2024lud,Wu:2019rog,Wu:2021caw}.

Besides the $P_{c}$ and $P_{cs}$ states, another two  pentaquark candidates, $\Lambda_{c}(2910)$ and $\Lambda_{c}(2940)$, have also been discovered experimentally. In 2006, $\Lambda_{c}(2940)$ was observed in the $D^{0} p$ invariant mass spectra by the BABAR Collaboration~\cite{BaBar:2006itc}. Then, the Belle Collaboration confirmed the existence of $\Lambda_{c}(2940)$ in the $\Sigma_{c}(2455)^{0,++} \pi^{+,-}$ invariant mass distributions, using a 533 $\mathrm{fb}^{-1}$ data sample recorded by the Belle detector at or 60 $\mathrm{MeV}$ below the $\Upsilon (4S)$ resonance~\cite{Belle:2006xni}. In 2022, based on $772 \times 10^{6}$ $B\bar{B}$ events collected with the Belle detector at the $\mathrm{KEKB}$ asymmetric-energy $e^{+}e^{-}$ collider, the Belle Collaboration observed a new structure, $\Lambda_{c}(2910)$, in the $\Sigma_{c}(2455)^{0,++} \pi^{+,-}$ invariant mass spectrum of the $\bar{B}^{0} \rightarrow \Sigma_{c}(2455)^{0,++} \pi^{+,-} \bar{p}$ processes~\cite{Belle:2022hnm}. The PDG average of masses and widths of $\Lambda_{c}(2910)$ and $\Lambda_{c}(2940)$ are~\cite{ParticleDataGroup:2024cfk},
\begin{eqnarray}
\Lambda^{+}_{c}(2910) : &\mathrm{M}&=\left(2914 \pm 7 \right)\; \mathrm{MeV}, \nonumber\\ &\Gamma&=\left(52 \pm 27\right) \; \mathrm{MeV}, \nonumber\\ \Lambda^{+}_{c}(2940) : &\mathrm{M}&=\left(2939.6^{+1.3}_{-1.5}\right) \; \mathrm{MeV}, \nonumber\\ &\Gamma&=\left(20^{+6}_{-5}\right) \; \mathrm{MeV}, \label{Eq.1}
\end{eqnarray}
respectively.

Similar to the case of $P_c$ states, the observed masses of $\Lambda_{c}(2910)$ and $\Lambda_{c}(2940)$ are close to the threshold of $D^{\ast} N$, indicating that these two states can be regarded as the $D^{\ast} N$ pentaquark molecular states. In Ref.~\cite{Zhang:2012jk}, the author investigated the properties of $\Lambda_{c}(2940)$ by using the QCD sum rules, indicating that $\Lambda_{c}(2940)$ could be regarded as the $S-$wave $D^{\ast} N$ bound state with $J^{P}=3/2^{-}$. Moreover, the authors in Ref.~\cite{Xin:2023gkf} proposed that $\Lambda_{c}(2910)/\Lambda_{c}(2940)$ could be regarded as $D^{\ast} N$ molecular state with $J^{P}=3/2^{-}$ by using the QCD sum rules. Using the QDCSM model, the authors in Ref.~\cite{Yan:2022nxp} proposed that $\Lambda_{c}(2940)$ is likely to be interpreted as $D^{\ast} N$ molecular state with $J^{P}=3/2^{-}$. By considering $\Lambda_{c}(2940)$ as $D^{\ast}N$ molecular state with $J^{P}=1/2^{\pm}$ and $3/2^{-}$, respectively, the authors in Refs.~\cite{Dong:2009tg,Dong:2010xv,Dong:2011ys} estimated the two-body decay channels, $D^{0} p$, $\Sigma^{++}_{c} \pi^{-}$, $\Sigma^{0}_{c} \pi^{+}$ and three-body decay channels, $\Lambda_{c}(2286)^{+} \pi^{+} \pi^{-}$, $\Lambda_{c}(2286)^{+} \pi^{0} \pi^{0}$. In Ref.~\cite{Yue:2024paz}, the authors studied the decay properties of $\Lambda_{c}(2910)$ and $\Lambda_{c}(2940)$ by using an effective Lagrangian approach, and the results suggested that $\Lambda_{c}(2910)$ and $\Lambda_{c}(2940)$ could be assigned as $D^{\ast} N$ molecular states with $J^{P}=1/2^{-}$ and $J^{P}=3/2^{-}$, respectively, while the branching fractions for $ND$, $\Sigma_{c}\pi$, and $\Sigma^{\ast}_{c}\pi$ channels of $\Lambda_{c}(2910)$ and $\Lambda_{c}(2940)$ were estimated. More recently, the authors in Ref.~\cite{Shen:2025akt} investigated the $DN$-$D^{\ast}N$ coupled-channel interactions using both heavy quark effective theory and a flavor-symmetry-constrained effective Lagrangian approach. Their estimations indicated  that the mass spectrum of $S$-wave $D^{\ast}N$ systems can be matched to three $\Lambda_{c}$ exotic states. Among them, the masses of $\Lambda_{c}(2910)$ and $\Lambda_{c}(2940)$ align with those of $D^{\ast}N$ bound states with $J^{P}=1/2^{-}$ and $J^{P}=3/2^{-}$, respectively. In conclusion, both the spectrum estimations and the decay properties investigations support that $\Lambda_{c}(2910)$ and $\Lambda_{c}(2940)$ can be identified as $D^{\ast}N$ bound states with $J^{P}=1/2^{-}$ and $J^{P}=3/2^{-}$, respectively.

In addition to the mass spectrum and decay properties, we also notice some theoretical researches on the productions of $\Lambda_{c}(2910)$ and $\Lambda_{c}(2940)$. The authors in Ref.~\cite{Wang:2015rda} investigated the production of $\Lambda_{c}(2940)$ in the $\gamma n \rightarrow D^{-} \Lambda_{c}(2940)$ process, where $\Lambda_{c}(2940)$ was regarded as $D^{\ast 0}p$ molecular state with $J^{P}=1/2^{\pm}$, and the estimations showed that the dominant contributions come from the $t$-channel with $D^{\ast}$ exchange. Then, by considering $\Lambda_{c}(2940)$ as $D^{\ast 0} p$ molecular state with $J^{P}=1/2^{\pm}$, the authors in Ref.~\cite{Xie:2015zga} investigated the production of $\Lambda_{c}(2940)$ in the $\pi^{-}p \rightarrow D^{-} D^{0} p$ process by including the $s, u$, and $t$-channels, indicating that the total cross sections are of orders of $10$ $\mu$b at $P_{\pi}=15$ GeV. Moreover, the authors in Ref.~\cite{Guo:2025efg} studied the productions of $\Lambda_{c}(2910)$ and $\Lambda_{c}(2940)$ in the process $\pi p\to D^- D^0 p$, where both $\Lambda_{c}(2910)$ and $\Lambda_{c}(2940)$ are considered as $D^{\ast}N$ molecular states, and the estimations indicated that the dominant contributions to the cross sections come from the intermediate state $\Lambda_{c}(2910)$.

In terms of experimental conditions, the $\mathrm{\bar{P}ANDA}$ Collaboration proposed to search for the charmed baryons in the $p \bar{p}$ scattering process with the momentum range of $P_{\bar{p}}=[1.5,15]$ GeV~\cite{PANDA:2009yku}. Based on the high energy beam and the potential higher-energy configurations with the future upgrades, one can investigate the productions of $\Lambda_{c}(2910)/\Lambda_c(2940)$ in the $p \bar{p}$ scattering process. It is worth noting that the estimations in Ref.~\cite{Yue:2024paz, Shen:2025akt} suggested that both $\Lambda_{c}(2910)$ and $\Lambda_{c}(2940)$ could be considered as $D^{\ast}N$ molecular states with the $J^{P}$ quantum numbers to be $1/2^{-}$ and $3/2^{-}$, respectively. Thus, we propose to investigate the $p \bar{p} \rightarrow \bar{\Lambda}_{c} D^{0} p$ process with the contributions from both $\Lambda_{c}(2910)$ and $\Lambda_{c}(2940)$, while the light-meson exchange is also considered in the present work. Specifically, we first estimate the cross sections for the $p \bar{p} \rightarrow \bar{\Lambda}_{c} D^{0} p$ process. Then, the $D^{0} p$ invariant mass distributions and Dalitz plot are estimated to check the individual contributions of $\Lambda_{c}(2910)$ and $\Lambda_{c}(2940)$.

This work is organized as follows. After introduction, we present our estimations of the cross sections for $p \bar{p} \rightarrow \bar{\Lambda}_{c} D^{0} p$. In Section \ref{sec:MA}, the numerical results and relevant discussions of the cross sections and differential cross sections are presented. The last section is devoted to a short summary. \\

\section{$\Lambda_{c}(2910)$ and $\Lambda_{c}(2940)$ productions in the $p \bar{p}$ scattering process}
\label{sec:MS}
In the present estimations, we consider that both $\Lambda_{c}(2910)$ and $\Lambda_{c}(2940)$ are  $D^{\ast}N$ molecular states with the $J^{P}$ quantum numbers to be $1/2^{-}$ and $3/2^{-}$, respectively. In Fig.~\ref{Fig.1}, we present the Feynman diagrams for the $p \bar{p}$ scattering process, where diagrams (a) and (b) correspond to the $t$-channel light-meson and $D/D^{\ast}$ exchanges,  respectively. In the present calculations, we employ the effective Lagrangian approach to depict the relevant hadron interaction vertices. The effective Lagrangians for $NN\pi$, $NN\eta$, $NN\sigma$, $NN\rho$, $NN\omega$ $\Lambda_{c} N D$, $\Lambda_{c} N D^{\ast}$, $\Lambda^{\ast}_{c1} N D$, and $\Lambda^{\ast}_{c2} N D$ can be written as~\cite{Dong:2009tg,Dong:2010xv,Dong:2011ys,Wang:2015rda,Xie:2015zga,Guo:2025efg,Janssen:1996kx,Wu:2007fc,Huang:2011as,Chen:2025qxh,Dong:2014ksa,Chen:2011xk,Okubo:1975sc,Jackson:2015dva, Machleidt:1987hj, Schutz:1998jx},
\begin{eqnarray}
\mathcal{L}_{N N \pi}&=&-i g_{NN \pi} \bar{N} \gamma_{5} \vec{\tau} \cdot \vec{\pi} N,\nonumber\\
%\end{eqnarray}
%\begin{eqnarray}
\mathcal{L}_{N N \eta}&=&-i g_{NN \eta} \bar{N} \gamma_{5}  \eta N,\nonumber\\
%\end{eqnarray}
%\begin{eqnarray}
\mathcal{L}_{N N \sigma}&=&-g_{NN \sigma} \bar{N} N  \sigma,\nonumber\\
%\end{eqnarray}
%\begin{eqnarray}
\mathcal{L}_{N N \rho}&=&-g_{NN \rho} \bar{N} \Big[\gamma_{\mu}- \frac{\kappa_{NN \rho}}{2m_{N}}\sigma^{\mu \nu} \partial_{\nu} \Big]  \vec{\tau} \cdot \vec{\rho}_{\mu} N,\nonumber\\
%\end{eqnarray}
%\begin{eqnarray}
\mathcal{L}_{N N \omega}&=&-g_{NN \omega} \bar{N} \Big[\gamma_{\mu}- \frac{\kappa_{NN \omega}}{2m_{N}}\sigma^{\mu \nu} \partial_{\nu} \Big]  \omega_{\mu} N,\nonumber\\
%\end{eqnarray}
%\begin{eqnarray}
\mathcal{L}_{\Lambda_{c} N D}&=& ig_{\Lambda_{c} N D} \bar{\Lambda}_{c} \gamma_{5} N D +h.c.,\nonumber\\ 
%\end{eqnarray}
%\begin{eqnarray}
\mathcal{L}_{\Lambda_{c} N D^{\ast}}&=& g_{\Lambda_{c} N D^{\ast}} \bar{\Lambda}_{c} \gamma^{\mu} N D^{\ast}_{\mu}+h.c.,\nonumber\\ 
%\end{eqnarray}
%\begin{eqnarray}
\mathcal{L}_{\Lambda_{c1}^\ast ND} &=& i g_{\Lambda_{c1}^\ast ND} \bar{\Lambda}_{c1}^\ast ND +h.c,\nonumber\\
\mathcal{L}_{\Lambda_{c2}^\ast ND} &=& \frac{g_{\Lambda_{c2}^\ast ND}}{m_\pi} \bar{\Lambda}_{c2}^{\ast \mu} \gamma^5 N \partial_\mu D +h.c..  \label{Eq.2}
\end{eqnarray}
Hereafter, $\Lambda_{c1}^\ast$ and $\Lambda_{c2}^\ast$ refer to $\Lambda_c(2910)$ and $\Lambda_c(2940)$, respectively. Then, due to that the $\Lambda_{c}(2910)$ and $\Lambda_{c}(2940)$ states are both considered as $D^{\ast}N$ molecular states, the effective Lagrangians of the molecular states coupling to their components are~\cite{Yue:2024paz},
\begin{eqnarray}
\mathcal{L}_{\Lambda_{c1}^\ast N D^{\ast}}&=& g_{\Lambda_{c1}^\ast N D^{\ast}} \bar{\Lambda}_{c1}^\ast \gamma^{\mu} \gamma_{5}  N D^{\ast}_{\mu}+h.c.,\nonumber\\ 
%\end{eqnarray}
%\begin{eqnarray}
\mathcal{L}_{\Lambda_{c2}^\ast N D^{\ast}}&=& g_{\Lambda_{c2}^{\ast} N D^{\ast}} \bar{\Lambda}_{c2}^{\ast \mu} N D^{\ast}_{\mu} +h.c.. \label{Eq.3}
\end{eqnarray}

With the above effective Lagrangians, one can obtain the amplitudes corresponding to the $p \bar{p} \rightarrow \bar{\Lambda}_{c} D^{0} p$ process, which are,

\begin{widetext}
\begin{eqnarray}
\mathcal{M}_{\pi, \: p}&=& \Big[\bar{u}(p_{5}) \Big(-i g_{N N \pi} \gamma_{5}\Big) u(p_{2}) \Big] S^{0}\Big(k_{1},m_{\pi}, \Gamma_{\pi} \Big) \Big[\bar{v}(p_{1}) \Big(-i g_{N N \pi} \gamma_{5} \Big) S^{1/2} \Big(k_{2},m_{p},\Gamma_{p} \Big) \Big(i g_{\Lambda_{c} N D} \gamma_{5} \Big) v(p_{3}) \Big] \nonumber\\ &\times& F \Big(k_{2},m_{p}, \Lambda_{r} \Big) \Big[F\Big(k_{1},m_{\pi},\Lambda_{r} \Big) \Big]^{2},\nonumber\\
%\end{eqnarray}
%\begin{eqnarray}
\mathcal{M}_{\eta, \: p}&=& \Big[\bar{u}(p_{5}) \Big(-i g_{N N \eta} \gamma_{5}\Big) u(p_{2}) \Big] S^{0}\Big(k_{1},m_{\eta}, \Gamma_{\eta} \Big) \Big[\bar{v}(p_{1}) \Big(-i g_{N N \eta} \gamma_{5} \Big) S^{1/2} \Big(k_{2},m_{p},\Gamma_{p} \Big) \Big(i g_{\Lambda_{c} N D} \gamma_{5} \Big) v(p_{3}) \Big] \nonumber\\ &\times& F \Big(k_{2},m_{p}, \Lambda_{r} \Big) \Big[F\Big(k_{1},m_{\eta},\Lambda_{r} \Big) \Big]^{2},\nonumber\\
%\end{eqnarray}
%\begin{eqnarray}
\mathcal{M}_{\sigma, \: p}&=& \Big[\bar{u}(p_{5}) \Big(-g_{N N \sigma} \Big) u(p_{2}) \Big] S^{0}\Big(k_{1},m_{\sigma}, \Gamma_{\sigma} \Big) \Big[\bar{v}(p_{1}) \Big(-g_{N N \sigma} \Big) S^{1/2} \Big(k_{2},m_{p},\Gamma_{p} \Big) \Big(i g_{\Lambda_{c} N D} \gamma_{5} \Big) v(p_{3}) \Big] \nonumber\\ &\times& F \Big(k_{2},m_{p}, \Lambda_{r} \Big) \Big[F\Big(k_{1},m_{\sigma},\Lambda_{r} \Big) \Big]^{2},\nonumber\\
%\end{eqnarray}
%\begin{eqnarray}
\mathcal{M}_{\rho, \: p}&=& \Big[\bar{u}(p_{5}) \Big(-g_{N N \rho} \Big[\gamma_{\mu}-\frac{\kappa_{N N \rho}}{2 m_{N}} \sigma^{\mu \nu} (i k^{\nu}_{1}) \Big] \Big) u(p_{2}) \Big] S^{1}_{\mu \alpha}\Big(k_{1},m_{\rho}, \Gamma_{\rho} \Big) \Big[\bar{v}(p_{1}) \Big(-g_{N N \rho} \Big[\gamma_{\alpha}-\frac{\kappa_{N N \rho}}{2 m_{N}} \sigma^{\alpha \beta} (i k^{\beta}_{1}) \Big] \Big) \nonumber\\ &\times& S^{1/2} \Big(k_{2},m_{p},\Gamma_{p} \Big) \Big(i g_{\Lambda_{c} N D} \gamma_{5} \Big) v(p_{3}) \Big] \times F \Big(k_{2},m_{p}, \Lambda_{r} \Big) \Big[F\Big(k_{1},m_{\rho},\Lambda_{r} \Big) \Big]^{2},\nonumber\\
%\end{eqnarray}
%\begin{eqnarray}
\mathcal{M}_{\omega, \: p}&=& \Big[\bar{u}(p_{5}) \Big(-g_{N N \omega} \Big[\gamma_{\mu}-\frac{\kappa_{N N \omega}}{2 m_{N}} \sigma^{\mu \nu} (i k^{\nu}_{1}) \Big] \Big) u(p_{2}) \Big] S^{1}_{\mu \alpha}\Big(k_{1},m_{\omega}, \Gamma_{\omega} \Big) \Big[\bar{v}(p_{1}) \Big(-g_{N N \omega} \Big[\gamma_{\alpha}-\frac{\kappa_{N N \omega}}{2 m_{N}} \sigma^{\alpha \beta} (i k^{\beta}_{1}) \Big] \Big) \nonumber\\ &\times& S^{1/2} \Big(k_{2},m_{p},\Gamma_{p} \Big) \Big(i g_{\Lambda_{c} N D} \gamma_{5} \Big) v(p_{3}) \Big] \times F \Big(k_{2},m_{p}, \Lambda_{r} \Big) \Big[F\Big(k_{1},m_{\omega},\Lambda_{r} \Big) \Big]^{2},\nonumber\\
%\end{eqnarray}
%\begin{eqnarray}
\mathcal{M}_{D, \: \Lambda_{c}}&=& \Big[\bar{u}(p_{5}) \Big(i g_{\Lambda_{c} N D} \gamma_{5} \Big) S^{1/2}\Big(k_{3},m_{\Lambda_{c}}, \Gamma_{\Lambda_{c}} \Big) \Big(ig_{\Lambda_{c} N D} \gamma_{5} \Big) u(p_{2}) \Big] S^{0}\Big(k_{1},m_{D},\Gamma_{D} \Big) \Big[\bar{v}(p_{1}) \Big(i g_{\Lambda_{c} N D} \gamma_{5} \Big) v(p_{3}) \Big] \nonumber\\ &\times& F \Big(k_{3},m_{\Lambda_{c}}, \Lambda_{r} \Big) \Big[F\Big(k_{1},m_{D},\Lambda_{r} \Big) \Big]^{2},\nonumber\\
%\end{eqnarray}
%\begin{eqnarray}
\mathcal{M}_{D^{\ast}, \: \Lambda_{c}}&=& \Big[\bar{u}(p_{5}) \Big(ig_{\Lambda_{c} N D} \gamma_{5} \Big) S^{1/2}\Big(k_{3},m_{\Lambda_{c}}, \Gamma_{\Lambda_{c}} \Big) \Big(g_{\Lambda_{c} N D^{\ast}} \gamma_{\mu} \Big) u(p_{2}) \Big] S^{1}_{\mu \nu}\Big(k_{1},m_{D^{\ast}},\Gamma_{D^{\ast}} \Big) \Big[\bar{v}(p_{1}) \Big(g_{\Lambda_{c} N D^{\ast}} \gamma_{\nu} \Big) v(p_{3}) \Big] \nonumber\\ &\times& F \Big(k_{3},m_{\Lambda_{c}}, \Lambda_{r} \Big) \Big[F\Big(k_{1},m_{D^{\ast}},\Lambda_{r} \Big) \Big]^{2},\nonumber\\
%\end{eqnarray}
%\begin{eqnarray}
\mathcal{M}_{D, \: \Lambda^{\ast}_{c1}}&=& \Big[\bar{u}(p_{5}) \Big(i g_{\Lambda^{\ast}_{c1} N D} \Big) S^{1/2}\Big(k_{3},m_{\Lambda^{\ast}_{c1}}, \Gamma_{\Lambda^{\ast}_{c1}} \Big) \Big(ig_{\Lambda^{\ast}_{c1} N D} \Big) u(p_{2}) \Big] S^{0}\Big(k_{1},m_{D},\Gamma_{D} \Big) \Big[\bar{v}(p_{1}) \Big(i g_{\Lambda_{c} N D} \gamma_{5} \Big) v(p_{3}) \Big] \nonumber\\ &\times& F \Big(k_{3},m_{\Lambda^{\ast}_{c1}}, \Lambda_{r} \Big) \Big[F\Big(k_{1},m_{D},\Lambda_{r} \Big) \Big]^{2},\nonumber\\
%\end{eqnarray}
%\begin{eqnarray}
\mathcal{M}_{D^{\ast}, \: \Lambda^{\ast}_{c1}}&=& \Big[\bar{u}(p_{5}) \Big(ig_{\Lambda^{\ast}_{c1} N D} \Big) S^{1/2}\Big(k_{3},m_{\Lambda^{\ast}_{c1}}, \Gamma_{\Lambda^{\ast}_{c1}} \Big) \Big(g_{\Lambda^{\ast}_{c1} N D^{\ast}} \gamma_{\mu} \gamma_{5} \Big) u(p_{2}) \Big] S^{1}_{\mu \nu}\Big(k_{1},m_{D^{\ast}},\Gamma_{D^{\ast}} \Big) \Big[\bar{v}(p_{1}) \Big(g_{\Lambda_{c} N D^{\ast}} \gamma_{\nu} \Big) v(p_{3}) \Big] \nonumber\\ &\times& F \Big(k_{3},m_{\Lambda^{\ast}_{c1}}, \Lambda_{r} \Big) \Big[F\Big(k_{1},m_{D^{\ast}},\Lambda_{r} \Big) \Big]^{2},\nonumber\\
%\end{eqnarray}
%\begin{eqnarray}
\mathcal{M}_{D, \: \Lambda^{\ast}_{c2}}&=& \Big[\bar{u}(p_{5}) \Big(\frac{ g_{\Lambda^{\ast}_{c2} N D}}{m_{\pi}} \gamma_{5} (-i p^{\mu}_{4}) \Big) S^{3/2}_{\mu \nu}\Big(k_{3},m_{\Lambda^{\ast}_{c2}}, \Gamma_{\Lambda^{\ast}_{c2}} \Big) \Big(\frac{g_{\Lambda^{\ast}_{c2} N D}}{m_{\pi}} \gamma_{5} (i k^{\nu}_{1}) \Big) u(p_{2}) \Big] S^{0}\Big(k_{1},m_{D},\Gamma_{D} \Big) \Big[\bar{v}(p_{1}) \Big(i g_{\Lambda_{c} N D} \gamma_{5} \Big) v(p_{3}) \Big] \nonumber\\ &\times& F \Big(k_{3},m_{\Lambda^{\ast}_{c2}}, \Lambda_{r} \Big) \Big[F\Big(k_{1},m_{D},\Lambda_{r} \Big) \Big]^{2},\nonumber\\
%\end{eqnarray}
%\begin{eqnarray}
\mathcal{M}_{D^{\ast}, \: \Lambda^{\ast}_{c2}}&=& \Big[\bar{u}(p_{5}) \Big(\frac{ g_{\Lambda^{\ast}_{c2} N D}}{m_{\pi}} \gamma_{5} (-i p^{\mu}_{4}) \Big) S^{3/2}_{\mu \nu}\Big(k_{3},m_{\Lambda^{\ast}_{c2}}, \Gamma_{\Lambda^{\ast}_{c2}} \Big) \Big(g_{\Lambda^{\ast}_{c2} N D^{\ast}} \Big) u(p_{2}) \Big] S^{1}_{\nu \theta}\Big(k_{1},m_{D^{\ast}}, \Gamma_{D^{\ast}} \Big) \Big[\bar{v}(p_{1}) \Big(g_{\Lambda_{c} N D^{\ast}} \gamma_{\theta} \Big) v(p_{3}) \Big] \nonumber\\ &\times& F \Big(k_{3},m_{\Lambda^{\ast}_{c2}}, \Lambda_{r} \Big) \Big[F\Big(k_{1},m_{D^{\ast}},\Lambda_{r} \Big) \Big]^{2},\nonumber\\
\end{eqnarray}
\label{Eq.5}
\end{widetext}
where the subscripts correspond to the propagators involved in the relevant processes. In addition, $\mathcal{S}^{0}(k_i,m_i,\Gamma_i)$ and $\mathcal{S}^{1}_{\mu \nu}(k_{i}, m_{i}, \Gamma_{i})$ are the propagators of scalar and vector meson with four momentum $k_i$, mass $m_i$, and width $\Gamma_i$, respectively, and the concrete forms are, 
\begin{eqnarray}
\mathcal{S}^{0}(k_i,m_i,\Gamma_i)&=&\frac{i} {k^2_{i}-m^2_{i} +i m_i \Gamma_{i}},\nonumber \\
%\end{eqnarray}
%\begin{eqnarray}
\mathcal{S}^{1}_{\mu \nu}(k_{i}, m_{i}, \Gamma_{i}) &=& \frac{-g^{\mu \nu}+k^{\mu}_{i} k^{\nu}_{i} / m^{2}_{i}}{k^{2}_{i}-m^{2}_{i}+i m_{i} \Gamma_{i}}.
\end{eqnarray}
In addition, $\mathcal{S}^{1/2}(k_{i},m_{i},\Gamma_{i})$ and $\mathcal{S}^{3/2}_{\mu \nu}(k_{i},m_{i},\Gamma_{i})$ are the propagators of $\Lambda_c(2910)$ and $\Lambda_c(2940)$, respectively, which read,
\begin{eqnarray}
&&\mathcal{S}^{1/2}(k_{i},m_{i},\Gamma_{i}) = \frac{\slash\!\!\!k_{i}+m_{i}} {k^2_{i}-m^2_{i} +i m_{i} \Gamma_{i}},\nonumber\\
%\end{eqnarray}
%\begin{eqnarray}
&&\mathcal{S}^{3/2}_{\mu \nu}(k_{i},m_{i},\Gamma_{i})  = \frac{\slash\!\!\!k_{i}+m_{i}} {k^2_{i}-m^2_{i} +i m_{i} \Gamma_{i}} \nonumber\\&&\qquad \qquad\times  \Big(-g^{\mu \nu} + \frac{\gamma^{\mu} \gamma^{\nu}}{3} + \frac{2 k^{\mu} k^{\nu}}{3 m^{2}_{i}} +\frac{\gamma^{\mu} k^{\nu}-k^{\mu} \gamma^{\nu}}{3 m_{i}} \Big).\qquad \label{Eq.7}
\end{eqnarray}

\begin{table}
\caption{The coupling constants involved in the $p \bar{p} \rightarrow \bar{\Lambda}_{c} D^{0} p$ process.}
\label{Tab:1}
\renewcommand{\arraystretch}{2}
\setlength{\tabcolsep}{5pt}
\centering
\begin{tabular}{cccc}
\hline\hline
Coupling constant&Value& Coupling constant&Value\\
\hline
$g_{\Lambda_{c} N D }$&$-13.98$&
$g_{\Lambda_{c} N D^{\ast} }$&$-5.20$\\
$g_{\Lambda^{\ast}_{c1} N D }$&$0.99$&
$g_{\Lambda^{\ast}_{c1} N D^{\ast} }$&$3.55$\\
$g_{\Lambda^{\ast}_{c2} N D}$&$0.84$&
$g_{\Lambda^{\ast}_{c2} N D^{\ast} }$&$2.34$\\
$g_{K D D^{\ast}_{s}}$&$5.0$&
$g_{K D^{\ast}_{s} D^{\ast}}$&$7.0$\\
$g_{K D_{s} D^{\ast}}$&$5.0$&
$g_{N N \pi}$&$13.46$\\
$g_{N N \eta}$&$4.76$&
$g_{N N \sigma}$&$9.42$\\
$g_{N N \rho}$&$3.25$&
$\kappa_{N N \rho}$&$6.10$\\
$g_{N N \omega}$&$11.76$&
$\kappa_{N N \omega}$&$0$\\
\hline \hline
\end{tabular} 
\end{table}

%%%%%%%%%%%%%%%%%%%%%%
\begin{figure*}[htbp]
     \includegraphics[width=170mm]{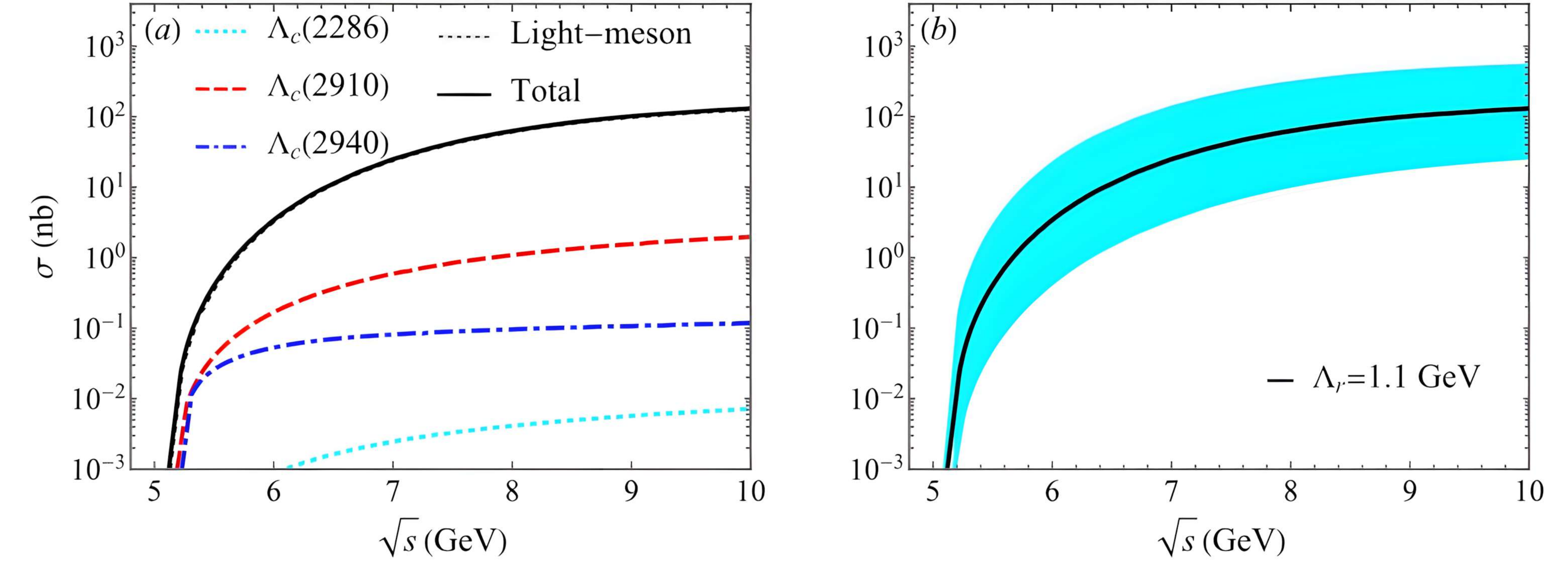}
     \caption{(Color online.) The cross sections for the $p \bar{p} \rightarrow \bar{\Lambda}_{c} D^{0} p$ process depending on the center-of-mass energy $\sqrt{s}$. Diagram ($a$) corresponds to the individual contributions from light-meson and $\Lambda^{(\ast)}_{c}$ states, while Diagram ($b$) corresponds to the total cross sections with the uncertainty resulted from the model parameter $\Lambda_{r}$, where the black solid curve is obtained with $\Lambda_{r}=1.1$ $\mathrm{GeV}$, and the cyan band represents the uncertainties resulted from the variation of $\Lambda_{r}$ from 1.0 to 1.2 $\mathrm{GeV}$.} 
     \label{Fig.2}
\end{figure*}
%%%%%%%%%%%%%%%%%%%%
%%%%%%%%%%%%%%%%%%%%%%
\begin{figure*}[htbp]
\includegraphics[width=170mm]{dcs_ppbar-new.pdf}
\caption{(Color online.) The Dalitz plot  for $p \bar{p} \rightarrow \bar{\Lambda}_{c} D^{0} p$ process at $\sqrt{s}=8$ $\mathrm{GeV}$ with $\Lambda_r=1.1 $ GeV depending on $m_{D^0 p}$ and $m_{\bar{\Lambda}_c D^0}$ (diagram ($b$)). Diagrams ($a$) and ($c$) correspond to the $\bar{\Lambda}_{c} D^{0}$ and $D^{0}p$ invariant mass distributions, respectively.}
     \label{Fig.3}
\end{figure*}
%%%%%%%%%%%%%%%%%%%%

In the above amplitudes, the form factor $F (k_{i},m_{i}, \Lambda_{r})$ is introduced to depict the inner structure of the involved hadrons and the off-shell effects, and its specific expression is,
\begin{eqnarray}
F(k_{i},m_{i},\Lambda_{r})
&=&\frac{\Lambda^{4}_{r}} {\Lambda^{4} _{r} +(k^{2}_{i}-m^{2}_{i})^2},\label{Eq.8}
\end{eqnarray}
where $k_{i}$ and $m_{i}$ are the four momentum and the mass of the exchanged state, respectively. It should be noted that the above form factor is typically applied to $s$-channel processes. In Refs. \cite{Janssen:2001wk, Shklyar:2005xg, Xie:2013db}, such kind of form factor have been extended to apply to $u$- and $t$- channels, which plays the role that smoothly suppresses the contributions from large momentum transfers while maintaining the physical behavior of the low energy region. In the above form factor, a model parameter $\Lambda_r$ is introduced	, and its specific value should be determined by comparing the theoretical estimations with the corresponding experimental measurements. However, no experimental measurement for the considered process is available at present. In our previous work, when we evaluated the cross sections for the $\pi^- p \to D^{\ast-} \Lambda_c$ in the effective Lagrangian approach, we find that the cross sections at $P_{\pi}=13$ GeV is estimated to be $2.59_{-2.00}^{+7.15}$ nb when taking $\Lambda_r=1.0\sim 1.2$ GeV~\cite{Guo:2025efg}, which is consistent with the upper limit of the experimental data reported by AGS Collaboration at Brookhaven National Laboratory~\cite{Christenson:1985ms}. Thus in the present work, we take the same parameter range as that employed in the estimations of the processes $\pi^- p \to D^- D^0 p$ to estimate the cross sections for $p\bar{p} \to \Lambda_c D^0 p$ process.

\section{NUMERICAL RESULTS AND DISCUSSIONS}
\label{sec:MA}
\subsection{Coupling Constants}
Before the estimations of the cross sections for the $p \bar{p} \rightarrow \bar{\Lambda}_{c} D^{0} p$ process, the values of  relevant coupling constants should be clarified. In the present estimations, both $\Lambda_c(2910)$ and $\Lambda_c(2940)$ are considered as $D^\ast N$ molecular states. Therefore, the coupling constants relevant to $\Lambda_c(2910)/\Lambda_c(2940)$ and their components $D^\ast N$  could be determined by the compositeness condition~\cite{Weinberg:1965zz,Baru:2003qq,Lin:2017mtz}, which is
\begin{eqnarray}
g_{\Lambda^{\ast}_{c} N D^{\ast}}^{2}
=\frac{4 \pi} {4 M_{\Lambda^{\ast}_{c}} m_{N}} \frac{(m_{D^{\ast}} + m_{N})^{5/2}} {(m_{D^{\ast}} m_{N})^{1/2}} \sqrt{32 \epsilon},
\end{eqnarray}
where $\epsilon = m_{N} + m_{D^{\ast}} - m_{\Lambda^{\ast}_{c}}$ stands for the binding energy of $\Lambda^{\ast}_{c}$, and we take $\epsilon = 32$ $\mathrm{MeV}$ and 6.2 $\mathrm{MeV}$ for $\Lambda_{c} (2910)$ and $\Lambda_{c}(2940)$, respectively. In addition, the factor $1/(4 M_{\Lambda^{\ast}_{c}} m_{N})$ is introduced for the normalization of two involved fermion fields. The specific values of $g_{\Lambda_{c1}^\ast ND^{\ast}}$ and $g_{\Lambda_{c2}^\ast ND^{\ast}}$ estimated with the above relation are presented in Table~\ref{Tab:1}. 

For the coupling constants $g_{\Lambda_{c1}^\ast ND}$ and $g_{\Lambda_{c2}^\ast ND}$, using the effective Lagrangians in Eq.~\eqref{Eq.2}, one can obtain the corresponding two-body amplitudes $\mathcal{M}_{\Lambda^{\ast}_{c} \rightarrow N D}$. Then, the decay widths of the $\Lambda^{\ast}_{c} \rightarrow N D$ processes can be written as,
\begin{eqnarray}
\Gamma_{\Lambda^{\ast}_{c} \rightarrow N D} = \frac{1}{(2J+1)8\pi} \frac{|\vec{k}_f|}{M^{2}} \overline{|\mathcal{M}_{\Lambda^{\ast}_{c} \rightarrow N D}|^2}, \label{Eq.9}
\end{eqnarray}
where $M$ and $J$ are the mass and angular momentum of the initial $\Lambda^{\ast}_{c}$ states. $\vec{k}_f$ is the three momentum of the final states in the initial rest frame. It is worth noting that the decay properties of $\Lambda_c(2910)$ and $\Lambda_c(2940)$ have been investigated in Ref.~\cite{Yue:2024paz}, and the branching fractions of $ND$ channel for $\Lambda_c(2910)/\Lambda_c(2940)$ are estimated to be,
\begin{eqnarray}
&&\mathcal{B}(\Lambda_{c}(2910) \to N D) \simeq 40 \%,\nonumber\\ 
&&\mathcal{B}(\Lambda_{c}((2940)) \to N D) \simeq 11 \%, \label{Eq.10}
\end{eqnarray}
With the above branching fractions and the central values of the widths of $\Lambda_c(2940)$ and $\Lambda_c(2910)$ in Eq.~\eqref{Eq.1}, one can obtain the coupling constants $g_{\Lambda_{c1}^\ast ND}$ and $g_{\Lambda_{c2}^\ast ND}$, which are listed in Table~\ref{Tab:1}.

The coupling constants $g_{N N \pi}$, $g_{N N \eta}$, $g_{N N \sigma}$, $g_{N N \rho}$, $\kappa_{N N \rho}$, $g_{N N \omega}$, and $\kappa_{N N \omega}$ are obtained from Refs.~\cite{Machleidt:1987hj, Chen:2025qxh, Schutz:1998jx, Huang:2011as, Wu:2007fc}. The specific values of the coupling constants mentioned above are also listed in Table~\ref{Tab:1}.

\subsection{Cross Sections for the $p \bar{p}$ scattering process}

After clarifying the amplitudes and relevant coupling constants, the cross sections for the $p \bar{p} \rightarrow \bar{\Lambda}_{c} D^{0} p$ process can be estimated, and the total amplitude is,
\begin{eqnarray}
\mathcal{M}_{\mathrm{Tot}}&=&\mathcal{M}_{\pi, \: p}+\mathcal{M}_{\eta, \: p}+\mathcal{M}_{\sigma, \: p}+\mathcal{M}_{\rho, \: p}+\mathcal{M}_{\omega, \: p} \nonumber\\ &+& \mathcal{M}_{D, \: \Lambda_{c}}+\mathcal{M}_{D^{\ast}, \: \Lambda_{c}}+\mathcal{M}_{D, \: \Lambda^{\ast}_{c1}}+\mathcal{M}_{D^{\ast}, \: \Lambda^{\ast}_{c1}} \nonumber\\ &+&\mathcal{M}_{D, \: \Lambda^{\ast}_{c2}}+\mathcal{M}_{D^{\ast}, \: \Lambda^{\ast}_{c2}}.
\end{eqnarray}
With the above total amplitude, one can obtain the differential cross sections of the $p \bar{p} \rightarrow \bar{\Lambda}_{c} D^{0} p$ process, which is,
\begin{eqnarray}
d{\sigma}=\frac{1} {8(2 \pi)^4} \frac{1} {	\Phi} \overline{\left| {\mathcal{M}_{\mathrm{Tot}}}\right|^2 }d p^0 _5 d p^0 _3 d\cos\theta d \eta, \label{Eq.13}
\end{eqnarray}
where the flux factor $\Phi=4|{\vec{p_1}}|\sqrt{s}$, $\vec{p_1}$ represents the three momentum of the initial antiproton. $\sqrt{s}$ is the center-of-mass energy. $p^{0}_{3}$ and $p^{0}_{5}$ refer to the energy of outgoing $\bar{\Lambda}_{c}$ and proton, respectively.

With the above preparations, the cross sections for $p \bar{p} \rightarrow \bar{\Lambda}_{c} D^{0} p$ depending on the center-of-mass energy $\sqrt{s}$ are presented in Fig.~\ref{Fig.2}. Among them, Fig.~\ref{Fig.2}-$(a)$ corresponds to the cross section for $p \bar{p} \rightarrow \bar{\Lambda}_{c} D^{0} p$ estimated with $\Lambda_r=1.1$ GeV, where the cyan dotted, red dashed, blue dash-dotted, and black dotted curves stand for the individual contributions from $\Lambda_c(2286)$, $\Lambda_c(2910)$, $\Lambda_c(2940)$, and light-meson, respectively, while the black solid curve refers to the total cross sections for the $p \bar{p}\to \bar{\Lambda}_{c} D^{0} p$ process. From this figure, one can find that the contributions from $\Lambda_{c}(2286)$ are not shown when $\sqrt{s}<6$ GeV because its cross sections in this range are less than 1 pb. In addition, the cross section of $\Lambda_{c}(2910)$ is about 2 nb at $\sqrt{s}=10$ $\mathrm{GeV}$, which is about 250 times and 15 times of that from $\Lambda_{c}(2286)$ and $\Lambda_{c}(2940)$, respectively. Moreover, our estimations indicate that the cross section resulted from light-meson intermediate process is predominantly, which enhances the total cross section by more than one order of magnitude. In Fig.~\ref{Fig.2}-$(b)$, we present the total cross sections for the $p \bar{p} \rightarrow \bar{\Lambda}_{c} D^{0} p$ process, where the black solid curve is obtained with $\Lambda_{r}=1.1$ $\mathrm{GeV}$, while the cyan band represents the uncertainties resulted from the varying of $\Lambda_{r}$ from 1.0 to 1.2 $\mathrm{GeV}$. Our results show that the cross sections increase rapidly near the threshold, and then increase rather slowly with the increasing of $\sqrt{s}$. In particular, the total cross sections are estimated to be $(130.7^{+397.1}_{-103.9})$ nb at $\sqrt{s}=10$ $\mathrm{GeV}$. 

In addition to the cross sections, we present the Dalitz plot for $p \bar{p} \rightarrow \bar{\Lambda}_{c} D^{0} p$ process at $\sqrt{s}=8$ $\mathrm{GeV}$ with $\Lambda_r=1.1$ GeV depending on $m_{D^0 p}$ and $m_{\bar{\Lambda}_c D^0}$ (diagram ($b$)), the projections depending on  $m_{D^0 p}$ (diagram ($c$)) and $m_{\bar{\Lambda}_c D^0}$ (diagram ($a$)) in Fig.~\ref{Fig.3}. From the Dalitz plot, one can find clear enhancement near the $D^0 p$ threshold.  In the $\bar{\Lambda}_{c} D^{0}$ and $D^{0}p$ invariant mass distributions, the cyan dotted, red dashed, blue dash-dotted, and black dotted curves correspond to the individual contributions from $\Lambda_c(2286)$, $\Lambda_c(2910)$, $\Lambda_c(2940)$, and light-meson, respectively, while the black solid curve refers to the summation of the individual contributions as well as their interferences. In Fig.~\ref{Fig.3}-$(a)$, our estimations indicate that the $\bar{\Lambda}_{c} D^{0}$ invariant mass spectrum is smooth across the entire mass range and shows no sharp structure. In Fig.~\ref{Fig.3}-$(c)$, our estimations show that the contribution from $\Lambda_c(2286)$ is relatively small, while the contribution from light-meson is rather large and smooth, which play the role of background of the observations of $\Lambda_c(2910)/\Lambda_c(2940)$. As for $\Lambda_c(2910)$ and $\Lambda_c(2940)$, our estimations suggested that the signal of $\Lambda_c(2910)$ is predominant, which is about one order larger than that of $\Lambda_c(2940)$. In addition, as shown in the inner figure of Fig.~\ref{Fig.3}-(c), a clear structure can be observed between 2.9 and 3.0 GeV in the $D^{0}p$ invariant mass spectra, which should corresponds to $\Lambda_c(2910)$ rather than $\Lambda_c(2940)$. 

%$ \vspace{0.2cm}$

\section{SUMMARY}
The observed masses of $\Lambda_{c}(2910)$ and $\Lambda_{c}(2940)$ are close to the threshold of $D^{\ast} N$, and the mass splitting of $\Lambda_{c}(2910)$ and $\Lambda_{c}(2940)$ is similar to the pentaquark candidates $P_{c}(4440)$ and $P_{c}(4457)$. These particular properties suggest that both $\Lambda_{c}(2910)$ and $\Lambda_{c}(2940)$ are good candidates of the $D^{\ast} N$ pentaquark molecular states with different $J^P$ quantum numbers. In addition to the decay behaviors, the production studies of $\Lambda_{c}(2910)$ and $\Lambda_{c}(2940)$ also provide useful information for the inner structures of these two $\Lambda_{c}$ states. Experimentally, the $\mathrm{\bar{P}ANDA}$ Collaboration proposed to search for the charmed baryons in the $p \bar{p}$ scattering process with $P_{\bar{p}}=[1.5,15]$ GeV. Therefore, we propose to investigate the $\Lambda_{c}(2910)$ and $\Lambda_{c}(2940)$ productions in the $p \bar{p} \rightarrow \bar{\Lambda}_{c} D^{0} p$ process in the present work.

For the individual contributions, our results show that the cross section of $\Lambda_{c}(2910)$ is about 2 nb at $\sqrt{s}=10$ $\mathrm{GeV}$, which is about 250 times and 15 times of that from $\Lambda_{c}(2286)$ and $\Lambda_{c}(2940)$, respectively. Moreover, our estimations indicate that the cross section resulted from light-meson intermediate process plays a significant role, which enhances the total cross section by more than one order of magnitude. In particular, the total cross sections for $p \bar{p} \rightarrow \bar{\Lambda}_{c} D^{0} p$ are estimated to be $(130.7^{+397.1}_{-103.9})$ nb at $\sqrt{s}=10$ $\mathrm{GeV}$, where the central value is estimated with the parameter $\Lambda_{r}=1.1$ GeV, and the uncertainties come from the variation of $\Lambda_{r}$ from 1.0 to 1.2 GeV.

The differential cross sections for the $p \bar{p} \rightarrow \bar{\Lambda}_{c} D^{0} p$ process are also estimated with $\sqrt{s}=8$ GeV in this work. For the $D^{0}p$ invariant mass spectrum, the contributions from $\Lambda_{c}(2286)$ and light-meson play the role of the background. In addition, our estimations indicate that the dominant contributions to the structure between 2.9 and 3.0 GeV come from $\Lambda_{c}(2910)$, indicating that the expected structure in the $D^{0}p$ invariant mass spectra should corresponds to $\Lambda_{c}(2910)$ rather than $\Lambda_{c}(2940)$. It is expected that the estimations in the present work could be tested by further experimental measurements at $\mathrm{\bar{P}ANDA}$ in future.

%$\hspace{1cm}$

\section*{ACKNOWLEDGMENTS}
This work is partly supported by the National Natural Science Foundation of China under the Grant Nos. 12175037 and 12335001, as well as supported, in part, by National Key Research and Development Program under the contract No. 2024YFA1610503. Quan-Yun Guo is also supported by the SEU Innovation Capability Enhancement Plan for Doctoral Students (Grant No. CXJH SEU 26160).

\end{document}